%% file: main.tex
\documentclass[11pt,a4paper]{article}

\usepackage[T1]{fontenc}
\usepackage{lmodern}
\usepackage[margin=1in]{geometry}
\usepackage{graphicx}
\usepackage{microtype}
\usepackage{hyperref}

\hypersetup{
  colorlinks=true,
  linkcolor=blue,
  citecolor=blue,
  urlcolor=blue,
  pdftitle={Celebi: Towards Reproducible and FAIR Physics Analysis},
  pdfauthor={Mingrui Zhao}
}

\title{Celebi: Towards Reproducible and FAIR Physics Analysis}
\author{%
  Mingrui Zhao\thanks{%
    \href{mailto:mingrui.zhao@mail.labz0.org}{mingrui.zhao@mail.labz0.org};
    \href{https://orcid.org/0000-0002-2858-2167}{ORCID: 0000-0002-2858-2167}.}\\
  \normalsize School of Physics, State Key Laboratory of Nuclear Physics and Technology\\
  \normalsize Peking University, Beijing, China
}
\date{}

\begin{document}
\maketitle

\begin{abstract}
Modern physics analyses evolve continuously as datasets, selections,
parameters, software environments, and analysis strategies are updated.
Over the lifetime of an analysis, this evolution can make it difficult to
determine the conditions under which an earlier result was produced or to
transfer the analysis reliably between collaborators. I designed and
developed Celebi as an analysis-management framework that maintains a
persistent connection between evolving analysis materials and their
generated results. Celebi organizes reusable algorithms, tasks, and their
dependencies into a structured workflow, preserves selected states as
immutable impressions, and associates retained results with the states
responsible for their production. This organization supports analysis
modification, systematic variations, collaboration, and review while the
working analysis continues to evolve. Celebi supports controlled rerunning
of preserved analysis states. When results remain retained, they can be
inspected and compared together with their corresponding states. By
capturing provenance during everyday analysis development, Celebi reduces
the need to reconstruct analysis history retrospectively and provides a
practical foundation for more transparent and reusable physics analysis.
\end{abstract}

\input{manuscript}

\end{document}

%% file: manuscript.tex
\section{Introduction}

Open and reproducible research requires preserving not only scientific
results but also the computational processes through which they are
obtained~\cite{nature:monya,Chen:2018drk}.
This is particularly important in high-energy physics, where analyses
involve complex processing chains and evolve over long periods.
While large experiments maintain centralized data-processing workflows,
individual physics analyses involve changing selections, configurations,
statistical procedures, and systematic studies.
Managing these evolving analyses while retaining the context of their
results is the focus of this work.

As an analysis develops, changes to code, parameters, inputs, and
dependencies can make it difficult to recover the conditions under which
earlier results were produced.
For example, a result presented during an internal review may no longer
correspond to the current analysis after subsequent modifications.
This challenge becomes more pronounced in collaborative projects, where
multiple configurations and analysis variations must coexist.
Reliable analysis management therefore requires a persistent connection
between each generated result and the analysis state responsible for
its production.

Existing tools address different aspects of reproducibility and
preservation.
Snakemake~\cite{molder2021sustainable} manages computational dependencies
and workflow execution, while REANA~\cite{reana} supports reproducible
execution in containerized environments.
The CERN Analysis Preservation portal (CAP)~\cite{Chen:2016unr}
provides infrastructure for documenting and preserving analyses.
Maintaining the connection between an evolving working analysis and the
results produced throughout its development, however, remains a distinct
analysis-management challenge.

Celebi addresses this challenge by organizing analysis materials into
structured components and preserving selected analysis states.
Generated results remain associated with the states under which they
were produced, even as the working analysis continues to evolve.
By separating lightweight analysis definitions from generated data
products while retaining their relationships, Celebi facilitates
analysis modification, algorithm reuse, systematic variations, and
historical result management.

This paper focuses on the practical implications of Celebi for physics
analysis development, collaboration, review, and preservation.
Its underlying workflow architecture, repository organization, and
impression mechanism are described in a companion paper~\cite{Zhao:CelebiCHEP}.

\section{Overview of the Celebi system}
\label{sec:overview}

Celebi separates potentially large data products generated during
execution from the lightweight materials defining a physics analysis,
including code, configurations, parameters, documentation, and
dependencies.
Its two main components are the \emph{Analysis Repository}, which
maintains the evolving analysis definition, and the \emph{Production
Factory}, which handles execution and generated results.
The middleware layer \emph{Yuki} connects the two, allowing preserved
analysis states to be executed and their results to be associated with
the corresponding analysis definitions.

Celebi distinguishes the editable working analysis from immutable
\emph{impressions}.
An impression identifies a preserved state of an analysis component
together with the preserved states of its dependencies.
Results are associated with task impressions, so their provenance
remains identifiable even after the working analysis changes.
Celebi supports controlled rerunning of preserved analysis states.
Such rerunning depends on the continued availability of the required
inputs, software environments, and execution infrastructure.

\section{Celebi throughout the physics analysis lifecycle}
\label{sec:analysis-management}

Celebi simplifies the management of evolving physics analyses by
organizing analysis components and their dependencies into a structured,
modular representation.
Analysts can modify, reuse, and reorganize components while retaining
the relationships between preserved analysis configurations and their
results.

\subsection{Analysis organization and evolution}
\label{sec:analysis-evolution}

Changes to inputs, algorithms, or configurations may require several
downstream computations to be updated or repeated.
Celebi simplifies the management of such changes by explicitly
representing analysis components and their dependencies.
Each component records its own dependencies, removing the need for a
single, centrally maintained workflow description.
Together, these relationships define the workflow and allow affected
downstream tasks to be identified without manually tracing scripts or
intermediate files.
Individual components can thus be modified without rewriting unrelated
parts of the analysis.

Celebi also separates reusable algorithms from the tasks that apply
them.
An algorithm defines a computational procedure, while a task
specifies its application with particular parameters and upstream
dependencies.
The same algorithm can be used by multiple tasks without duplicating
its implementation, for example when fitting different datasets or
studying alternative configurations.

This modular organization facilitates the management of related
analysis configurations.
Because components are represented as structured repository objects,
they can be copied, reorganized, and modified through familiar
file-system-like operations while retaining their dependency
relationships.
Optimization studies and systematic variations can consequently share
unchanged components while modifying only the parts that differ.
This reduces duplication and makes differences between configurations
more explicit.

\subsection{Provenance-aware result management}
\label{sec:provenance-management}

Celebi manages generated results in association with the analysis
states responsible for their production, rather than treating them as
independent output files.
Results organized only by filenames or directory structures can become
difficult to interpret as the analysis evolves, particularly when the
original configuration is no longer the current one.

\begin{figure}[htbp]
    \centering
    \includegraphics[width=0.70\linewidth]{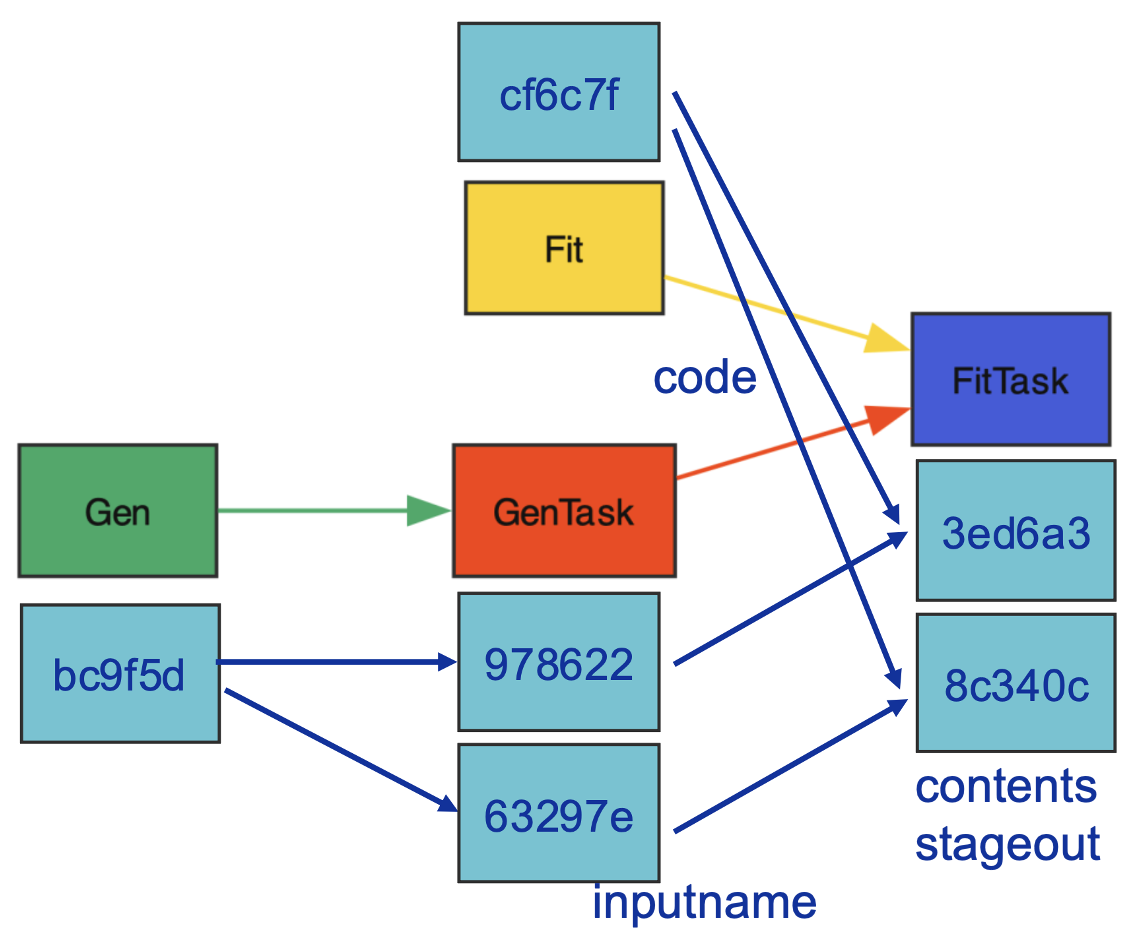}
    \caption{Illustration of dependency-aware impressions in Celebi.
    Analysis components and their preserved impressions are shown as
    separate objects; impression references identify the preserved
    upstream states associated with a task.}
    \label{fig:impressions}
\end{figure}

In Celebi, results are associated with immutable task impressions.
These impressions identify preserved code and configuration, including
task parameters, together with the exact impressions of upstream
dependencies. This allows the provenance of a result to be traced
through the analysis.
Even if a task uses the same algorithm and parameters, a change to an
upstream input can produce a distinct preserved analysis state.
Recording the dependency impressions makes it possible to distinguish
these configurations and keep their results associated with the
appropriate states.
Figure~\ref{fig:impressions} illustrates this relationship.

For example, a mass-fit result presented during an internal review
may later be superseded following changes to the event selection or
fitting model.
Recovering the earlier configuration would otherwise require manually
matching the result to the relevant code, parameters, and inputs.
When results remain retained, they can be accessed and compared together
with their respective preserved analysis states, making it easier to
identify differences in their configurations.

This approach complements conventional source-code version control.
While Git tracks repository-level changes, Celebi explicitly connects
results to preserved states of individual analysis components and
their dependencies, reducing the need for manually maintained
provenance records.

\subsection{Collaboration, review, and preservation}
\label{sec:collaboration}

Physics analyses often involve multiple collaborators and several
rounds of internal review.
Effective collaboration requires a shared understanding of analysis
structure, computational procedures, and the configurations used to
produce results.
When this information is scattered across scripts, configuration files,
documentation, and personal directories, an analysis can be difficult
for others to navigate and understand.

Celebi organizes analysis materials, documentation, and dependencies
into structured repository objects.
Collaborators can inspect a selected component together with its local
documentation and dependencies, without first reconstructing the
complete analysis from unrelated files.
This helps collaborators understand and contribute to specific parts
of an analysis.

\begin{figure}[htbp]
    \centering
    \includegraphics[width=0.88\linewidth]{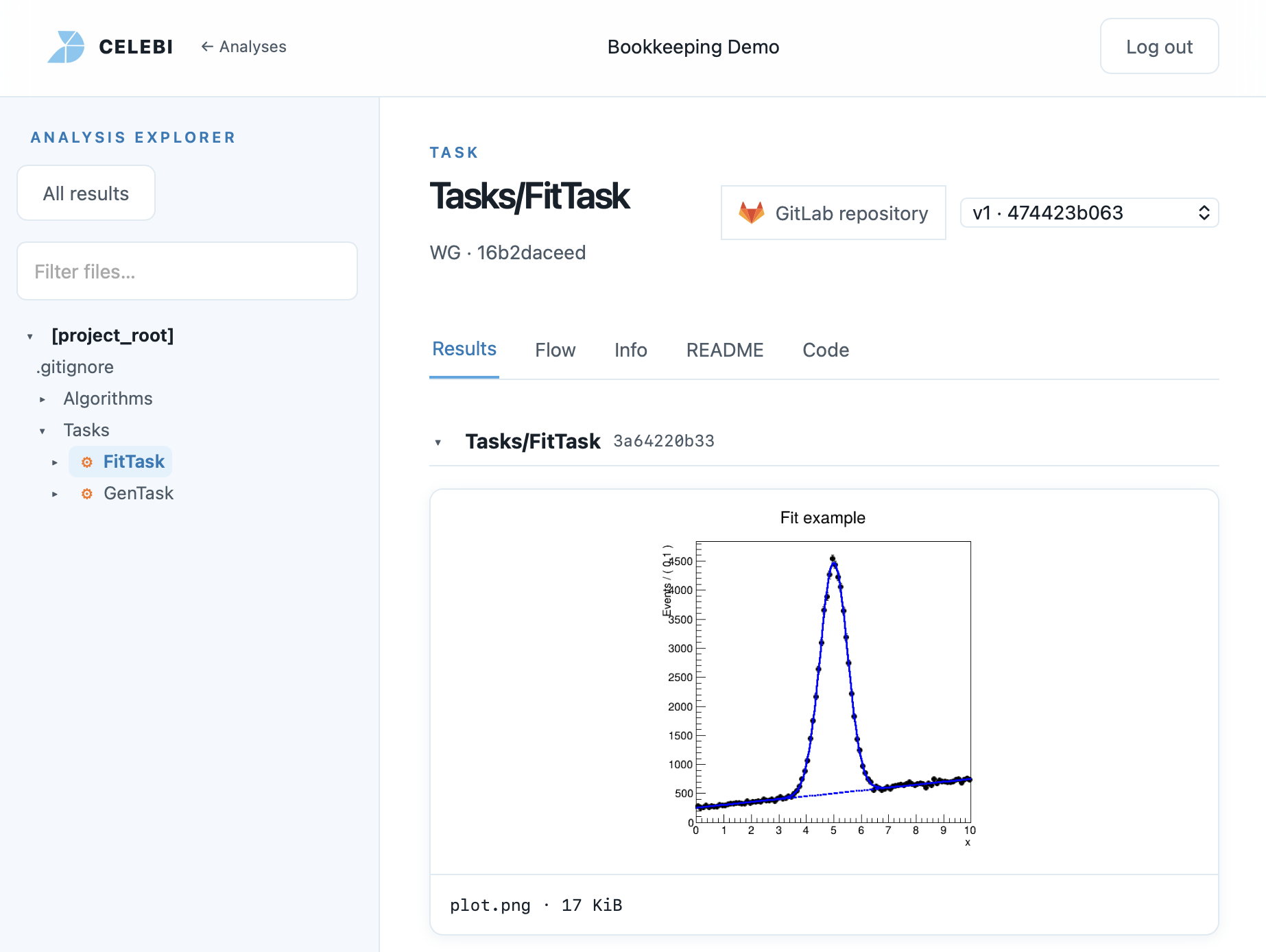}
    \caption{Celebi bookkeeping interface showing the analysis
    explorer, a fitting task, a selected preserved version, and its
    associated result. The interface also provides access to task
    information, documentation, and code for inspection and review.}
    \label{fig:bookkeeping}
\end{figure}

Immutable impressions also give collaborators and reviewers stable
references to particular analysis states.
Reviewers can inspect the configuration associated with a result even
after the working analysis has evolved.
When changes are introduced in response to review comments, updated
states and results can be retained alongside those examined in earlier
review rounds, facilitating comparisons between revisions.
Figure~\ref{fig:bookkeeping} shows the Celebi bookkeeping interface,
through which users can browse analysis components and inspect results
associated with a selected preserved state.

The same organization facilitates analysis handover and long-term
preservation.
Structured components and accompanying documentation reduce reliance
on individual conventions and personal knowledge when an analysis
is transferred to new collaborators.
By capturing provenance during ordinary analysis development, Celebi
reduces the need to reconstruct analysis history retrospectively when
results are reviewed, transferred, or preserved.

\subsection{Towards FAIR physics analysis}
\label{sec:fair}

These capabilities provide a basis for extending FAIR principles from
research data to the computational analyses that produce scientific
results.
Reusing a result requires not only its output data but also sufficient
information about the code, configurations, parameters, dependencies,
and provenance involved in its production.

By associating results with identifiable analysis states, Celebi
provides a foundation for finding and reusing analysis materials and
outputs.
Its structured representation supports programmatic interpretation
and can facilitate interoperability with other infrastructures, while
the separation of analysis definitions from generated data facilitates
their independent preservation.
These features do not by themselves establish full FAIR compliance,
but contribute to more transparent and reusable physics analyses.

\section{Summary}

Celebi simplifies the management of physics analyses throughout their
lifecycle.
Its modular organization and explicit dependencies make analysis
components easier to modify, reuse, and reorganize, while immutable
impressions retain the link between generated results and the analysis
states responsible for their production.
These capabilities support historical result inspection, collaboration,
and review with less reliance on manual bookkeeping. They also support
systematic variations and controlled rerunning of preserved states,
provided that the required inputs, software environments, and execution
infrastructure remain available.

Maintaining analysis states and provenance during everyday development
also provides a natural starting point for long-term preservation.
In this way, Celebi offers a practical foundation for extending FAIR
principles beyond research data to the computational analyses used to
produce scientific results.